\documentclass[aps,prl,reprint,showpacs,article,floatfix]{revtex4-1}
\pdfoutput=1 

\usepackage{fullpage}
\usepackage{amsmath}
\usepackage{amssymb}
\usepackage{graphicx}

\begin{document}

\title{Instability of 
Super--Entropic Black Holes in Extended Thermodynamics}
\author{Clifford V. Johnson}
\email{johnson1@usc.edu}
\affiliation{Department of Physics and Astronomy\\ University of
Southern California \\
 Los Angeles, CA 90089-0484, U.S.A.}


\begin{abstract}
The charged black hole of Ba\~nados, Teitelbiom and Zanelli is studied in extended gravitational thermodynamics where there is a dynamical pressure and volume. It is a simple  example of a super--entropic black hole,   violating the reverse isoperimetric inequality. It is proven that this property implies that its specific heat at constant volume is  negative, signalling a new kind of fundamental instability for black holes. It is  conjectured that this instability is present for other super--entropic black holes, and this is demonstrated numerically for a large family of known solutions. 
\end{abstract}


\maketitle

Combining quantum mechanics with spacetimes possessing horizons reveals black holes to be thermodynamic objects~\cite{Bekenstein:1973ur,Bekenstein:1974ax,Hawking:1974sw,Hawking:1976de}\footnote{For a recent historical overview, see {\it e.g.,} ref.~\cite{Grumiller:2014qma}.}. 
The standard  framework assigns to a black hole a temperature~$T$, an  entropy~$S{=}A/4$, and an energy $U{=}M$, which obey the first law of thermodynamics: $dU{=}TdS$.  (Here,~$M$ is the mass of the black hole,~$A$ is the area of its horizon, and we are using geometrical units where $G,c,\hbar,k_{\rm B}$ have been set to unity.) The most celebrated example of this is supplied by the Schwarzschild black hole in $d{=}4$ spacetime dimensions, in asymptotically flat spacetime, which has $S{=}4\pi M^2$ and $T{=}1/8\pi M$.  The black hole  reduces its mass--energy $M$ due to Hawking radiation at temperature $T$, but this shrinking results in higher temperature, increased radiation,  and hence more mass loss,~{\it etc}. This is an unstable situation, and is the earliest known quantum instability of a gravitational system. It is summarized neatly by observing that the specific heat is negative: 
\begin{equation}
\label{eq:hawking-explosion}
C=T\frac{\partial S}{\partial T} = -\frac{1}{8\pi\,T^2}\ .
\end{equation}
In this Letter,  a new but closely related quantum instability for black holes will be uncovered, in  a context that has been the subject of considerable research activity in recent years. The setting is {\it extended}  gravitational thermodynamics~\footnote{For a recent review see ref.~\cite{Kubiznak:2016qmn}.}. There,   making dynamical  the cosmological constant,~$\Lambda$,  of the gravity theory  yields a pressure variable  $p{=}{-}\Lambda/8\pi$,  and the black hole's mass $M$ is found~\cite{Kastor:2009wy} to be equal to the  enthalpy $H{=}U{+}pV$. The pressure's  conjugate is the thermodynamic volume~$V{\equiv}(\partial H/\partial p)_S$. 
The first law of thermodynamics is $dH{=}TdS{+}Vdp$. 
For spacetimes with  $\Lambda{<}0$,  considered here,   the pressure is positive, yielding a well--defined equilibrium thermodynamic framework. In such spacetimes, there are analogues of the Schwarzschild black holes  which come in two classes~\cite{York:1986it,Hawking:1982dh}: Those which are large compared to the length scale $\ell\,{\sim}(-\Lambda)^{-1/2}$  have  a $T$ that grows with $M$, which makes them stable. Those which are small compared to $\ell$ are the unstable analogues of flat--space Schwarzschild, and for those, in $d$ spacetime dimensions~\cite{Johnson:2015ekr}: 
\begin{eqnarray}
\label{eq:ceepeetee}
C_p(T)=- \frac{d-2}{4}\omega_{d-2}\left(\frac{d-3}{4\pi}\right)^{d-2}\frac{1}{T^{d-2}}+\cdots\ ,\quad 
\end{eqnarray}
(where $\omega_n{=}2\pi^{(n+1)/2}/\Gamma[(n+1)/2]$ is the volume of the round unit sphere $S^n$ {\it i.e.,} $\omega_1{=}2\pi, \omega_2{=}4\pi, \cdots$). In the above, the notation $C_p$ was used, indicating that since the specific heat usually discussed  in traditional thermodynamics is at a given $\Lambda$, it is at constant pressure, $p$. For $d{=}4$ the above result indeed reduces to that of equation~(\ref{eq:hawking-explosion}). In extended thermodynamics, it is (as emphasized recently in ref.~\cite{Johnson:2019vqf}) prudent to also examine the behaviour of the specific heat at constant volume, $C_V$, since it can give important extra information about the physics. $C_V(T)$ is in fact a powerful probe of the nature of the underlying degrees of freedom. It will be shown below that for a number of examples of the so--called ``super--entropic'' black holes~\cite{Hennigar:2014cfa}, $C_V{<}0$, showing that they are unstable in this extended framework. The meaning of this instability will be explained. 

Because of the explicit way the cosmological constant enters  into Einstein's field equations, and hence into the line element encoding the black hole solution,  questions of constant pressure are easier to formulate in this framework. In general, since the volume does not appear explicitly, it is harder to extract closed form expressions for quantities like~$C_V$ from the black hole equations. 
Nevertheless, progress can be made. First, on an exactly solvable example, and later by using numerical methods on a large class of examples. 

The black hole solution of Ba\~nados, Teitelbiom and Zanelli (BTZ)~\cite{Banados:1992wn,Banados:1992gq}  has played a central role in studies of the physics of black holes, both classical and quantum.  This is despite the fact that it  is $(2{+}1)$--dimensional, and that it has a negative cosmological constant,~$\Lambda$.  Its relative simplicity is part of the reason for its ubiquity, often providing simple (sometimes exact) results for a range of important phenomena from gravitational collapse to the scattering of quanta off the black hole.  It even plays a direct role in studies  of quantum aspects of classes of black holes in other dimensions, 
 organizing 
 the microscopic accounting of the quantum degrees of freedom that underlie entropy. (See refs.~\cite{Strominger:1997eq,Strominger:1996sh,Grumiller:2014qma}.)

Here, the BTZ black hole will again serve as a useful exact model,
but it is  the electrically charged version that will be of interest\footnote{The author thanks Felipe Rosso for bringing this solution to his attention.}. In fact, it is one of the simplest black hole solutions that has non--zero~$C_V$. It solves the  $(2{+}1)$--dimensional  Einstein--Maxwell system  with action:
\begin{equation}
I=-\frac{1}{16\pi }\int \! d^3x \sqrt{-g} \left(R-2\Lambda -F^2\right)\ ,
\label{eq:action}
\end{equation}
 where $\Lambda{=}{-}1/\ell^2$  sets a length scale $\ell$.  The metric and gauge potential are~\cite{Martinez:1999qi}:
\begin{eqnarray}
\label{eq:btz-black-hole}
ds^2 &=& -f( r)dt^2
+ f(r)^{-1}dr^2 + r^2 d\varphi^2\ ,\nonumber \\
f( r) &\equiv& -8M+\frac{Q^2}{2}\log\left({r}/{\ell}\right)+\frac{r^2}{\ell^2}\ , \nonumber\\
A_t &=& Q\log\left({r}/{\ell}\right)\ .
\end{eqnarray}
The  (circular) horizon is at $r{=}r_+$, the largest positive real root of $f(r_+){=}0$. The physical charge and mass are~$Q$ and  $M$, respectively. The extended thermodynamics was explored in ref.~\cite{Frassino:2015oca}, yielding:
\begin{eqnarray}
\label{eq:thermodynamic-quantities}
H&=&\frac{4pS^2}{\pi} - \frac{Q^2}{32}\log\left(\frac{32pS^2}{\pi}\right)\ ,\quad S=\frac{\pi}{2}r_+\ ,\nonumber\\
T&=&\frac{8pS}{\pi}-\frac{Q^2}{16S}\ ,\quad {\rm and}\quad V= \frac{4S^2}{\pi} -\frac{Q^2}{32p}\ .
\end{eqnarray}
A key  point here is that $V$ and $S$ are independent variables, and so $C_V{\neq}0$. The ultimate goal here is to compute its temperature dependence. It will be helpful to start with $C_p{\equiv}T\partial S/\partial T|_p$. For this, the expression for $T$ in equation~(\ref{eq:thermodynamic-quantities}) is readily inverted to  give (taking the plus branch, which has the correct $Q{=}0$ limit):
\begin{equation}
\label{eq:entropy-tee}
S=\frac{\pi T}{16p}\left[{1+\sqrt{1+2Q^2p/\pi T^2}}{}\right]\ ,
\end{equation}
resulting in:
\begin{eqnarray}
\label{eq:ceepee}
C_p(T)=\frac{\pi T}{16p}\left(1+\frac{1}{\sqrt{1+2Q^2p/\pi T^2}}\right)\ ,
\end{eqnarray}
which is manifestly positive, with the large $T$ behaviour $C_p(T){=}\pi T/8p+\cdots$.
Turning to $C_V(T)$, we can use either of the well--known  relations: 
\begin{equation}
\label{eq:specific-relation}
C_p-C_V=TV\alpha_p^2\kappa_T \ ,\qquad \frac{C_p}{C_V}=\frac{1}{\kappa_T\beta_S} \ ,
\end{equation}
where $\alpha_p{\equiv}V^{-1}\partial V/\partial T|_p$ is the isobaric thermal expansion coefficient, $\kappa_T{\equiv}{-}V\partial p/\partial V|_T$ is the isothermal bulk modulus, and $\beta_S{\equiv}{-}V^{-1}\partial V/\partial p|_S$ is the adiabatic compressibility.
The expression~(\ref{eq:entropy-tee}) for $S(T,p)$ can be used in the expression for the volume in equation~(\ref{eq:thermodynamic-quantities}) to yield:
\begin{equation}
\label{eq:veeteepee}
V(T,p)= \frac{\pi T^2}{64 p^2}\left[1+\sqrt{1+2Q^2p/\pi T^2}\right]^2 - \frac{Q^2}{32p}\  .
\end{equation}
From this, $\alpha_p$ and $\kappa_T$ can be readily computed (and the last of equation~(\ref{eq:thermodynamic-quantities}) can yield $\beta_S$), and either of equations~(\ref{eq:specific-relation}) yields:
\begin{eqnarray}
\label{eq:ceeveetee}
C_V(T) &=& -\frac{Q^2}{32 T }\times\\
&&\!\!\! \left[\frac{ 1+\sqrt{1+2Q^2p/\pi T^2}}{1+\sqrt{1+2Q^2p/\pi T^2}+3Q^2p/2\pi T^2}\right]\ , \nonumber
\end{eqnarray}
where  equation~(\ref{eq:veeteepee}) must be solved for $p(T,V)$ in order to write $C_V(T)$ explicitly in terms of the fixed~$V$, and some additional $T$ dependence. (For brevity, it is not written out here  as it is not needed.) This is the main result. The key observation here is that~$C_V$ is manifestly negative. Moreover, at large~$T$:
\begin{equation}
C_V(T)=-\frac{Q^2}{32\,T}+\cdots\ .
\end{equation}
This result shows that at fixed volume the  internal energy decreases if the temperature increases, or {\it vice--versa}:
\begin{equation}
U_{\rm f} - U_{\rm i}=-\frac{Q^2}{32}\ln(T_{\rm f}/T_{\rm i})\ .
\end{equation}
While the precise $T$--dependence of $C_V$ is different from $d{=}4$ Schwarzschild (because of dimensional analysis, and here $d{=}3$), (see equation~(\ref{eq:ceepeetee})) the phenomena being heralded are related. The black hole is unstable to shrinking, radiating away.  Recall that thermodynamic volume here is {\it not} geometric volume, so this is entirely consistent with fixed $V$. In fact, it is clear from equations~(\ref{eq:thermodynamic-quantities}) that moving up an isochore results in $T$ increasing and hence $p$ increasing. Increasing $p$ means decreasing $S$ if $V$ is to be constant. So the horizon area shrinks. The system radiates at higher temperature as a result, accelerating the evaporation process. This new type of instability is  possible because $p$ is now dynamical, and because the  volume here is decoupled from entropy and hence the black hole size.

The charged BTZ black hole is an example of a ``super--entropic'' black hole solution~\cite{Frassino:2015oca}. In other words, it violates the so--called ``reverse isoperimetric inequality" of ref.~\cite{Cvetic:2010jb}:
\begin{equation}
\label{eq:reverse-isoperimetric}
{\cal R} =\left(\frac{(d-1)V}{\omega_{d-2}}\right)^{\frac{1}{d-1}}\left(\frac{\omega_{d-2}}{A}\right)^{\frac{1}{d-2}} \ge1\ .
\end{equation} 
In the case of the Schwarzschild black hole, (or Reissner--Nordstr\"om), the horizon is simply a round sphere  with $A{=}\omega_{d-2}r_+^{d-2}$, and the thermodynamic volume is the naive geometric volume~\cite{Kastor:2009wy} $V{=}\omega_{d-2}r_+^{d-1}/(d{-}1)$. Then $S{=}A/4$ and $V$ are not independent and hence $C_V{=}0$.  In this case, the identity is saturated. For a large number of other solutions (including  Kerr black holes~\cite{Cvetic:2010jb}, STU black holes~\cite{Caceres:2015vsa}, Taub--NUT/Bolt~\cite{Johnson:2014xza}) where $C_V\neq0$, the volume factor in equation~(\ref{eq:reverse-isoperimetric}) is in fact larger, resulting in ${\cal R}>1$. So the identity essentially states that a black hole of a given thermodynamic volume has its   entropy bounded above by the  amount that a Schwarzschild black hole would have. To date, this has been simply an observation made in the literature, but no physical reason has been given for the bound. Solutions that violate the bound have been identified (the charged BTZ being the simplest), and in fact several have now been constructed and presented in the literature~\cite{Hennigar:2014cfa,Hennigar:2015cja}.

In fact, for the charged BTZ black hole it is easy to see that there is a direct connection between the violation of the identity and the $C_V{<}0$ instability under discussion here. An expression for $C_V$ in terms of $S$ and $V$ can be derived by first eliminating $p$ between the third and fourth members in  equations~(\ref{eq:thermodynamic-quantities}) to give $T(S,V)$:
\begin{equation}
\label{eq:teeseevee}
T=\frac{\pi V}{16S}\left(\frac{Q^2}{4S^2-\pi V}\right)\ .
\end{equation}
This is an interesting form, since the manifest positivity of the temperature, evident in other forms, is reflected in the statement that $4S^2{>}\pi V$. (Note that~$T$ is finite as $Q{\to}0$. The numerator and denominator go to zero at the same rate.) In fact, $4S^2{=}\pi V$ is exactly the saturation of the identity. Differentiating~(\ref{eq:teeseevee}) with respect to $T$, at fixed~$V$,  readily yields after some rearrangements\footnote{After this work was completed, it was noticed  that eq.~(\ref{eq:teeseevee}) and a rearrangement of eq.~(\ref{eq:ceevee}) were also derived in ref.~\cite{Mo:2017nhw}. Neither the sign of $C_V$, nor its significance, was noted there however.}:
\begin{equation}
\label{eq:ceevee}
C_V=-S\left(\frac{4S^2-\pi V}{12S^2-\pi V}\right)\ .
\end{equation}
In other words, the violation of the identity has ensured that $C_V{<}0$.

This result is highly suggestive and leads to a natural conjecture that super--entropic black holes always have $C_V{<}0$, making them unstable in extended thermodynamics. This is hard to test, because, as mentioned earlier, explicit expressions for~$C_V$ are hard to write, in general. It is reasonable to suppose however, that $C_V$, when written in terms of~$S$ and $V$ in general, will be a positive function for sub--entropic black holes, with a factor generalizing $(\pi V{-}4S^2)$  that goes negative for super--entropic holes.  (When $C_V(T)$ is positive, it is likely to have the Schottky--like peaks found in ref.~\cite{Johnson:2019vqf}.)

\begin{figure}[h]
\centering
\includegraphics[width=0.45\textwidth]{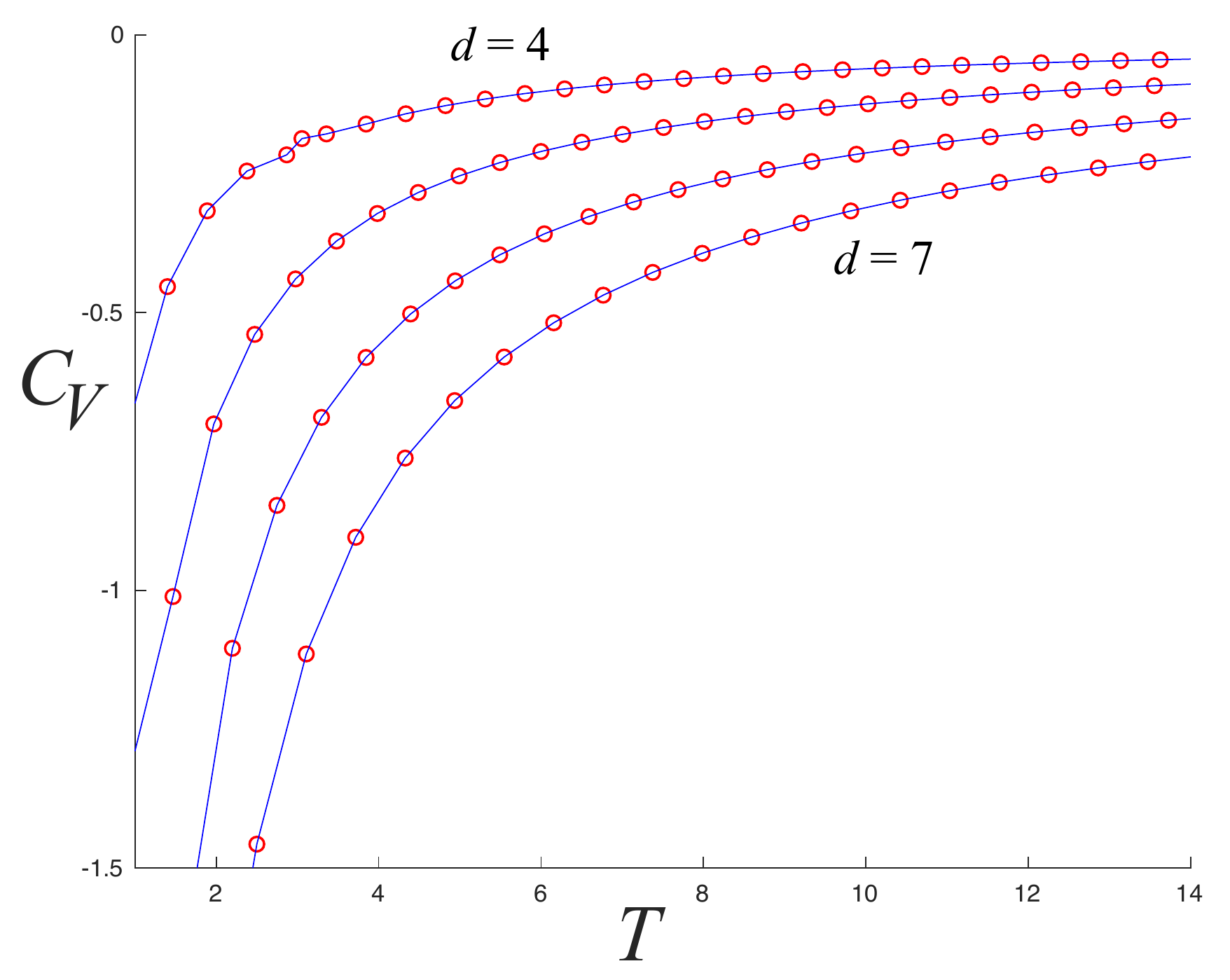}
\caption{\label{fig:cvplots} $C_V$ {\it vs.} $T$ at large $T$ for $d{=}4$ (top curve) to $d{=}7$ (bottom). $V\simeq56$ was used.}
\end{figure}

Although exact expressions for $C_V$ are difficult to extract, progress can still be made numerically, using the methods of ref.~\cite{Johnson:2019vqf}. As examples, the geometries that resulted from special  ultra--spinning limits of the Kerr black hole in $d$ spacetime dimensions in ref.~\cite{Hennigar:2014cfa} were tested. (See eq.~(25) of that paper for the explicit expressions.) Figure~\ref{fig:cvplots}
shows  examples of  the results for the large~$T$ behaviour of $C_V(T)$ for $d{=}4,5,6,7$. (Exploration showed that these are the tails of negative versions of the  kind of  peaks found in ref.~\cite{Johnson:2019vqf}.)
This large~$T$ behaviour, like the leading behaviour of $C_p(T)$ for the Schwarzschild--like black holes,  again signals a runaway evaporation. 

As a final remark that supports a connection between negative $C_V$ and super--entropicity, ref.~\cite{Johnson:2019vqf} emphasizes that $C_p(T)$ and $C_V(T)$ are two subtly different diagnostics of the available degrees of freedom of a system. They both measure the increase in energy when heat is added to the system, per unit~$T$ increase, but the former is larger because it must also take into account work done in increasing the volume. So $C_V$ measures only standard or ``traditional'' degrees of freedom that don't involve changes in volume to excite~\footnote{Ref.~\cite{Johnson:2019vqf} concludes  that this makes these gravitating systems in extended thermodynamics very different from familiar matter, despite  many observations in the literature about their similarity in phase structure~\cite{Kubiznak:2016qmn}. They have either zero or relatively few ``traditional'' degrees of freedom. This is exploited in ref.~\cite{Johnson:2019olt}.}. 
If there is an instability associated with some of these degrees of freedom ({\it i.e.,} if~$U$ can decrease on increasing $T$), then there must be an over--compensation in the  heat needed in order to have the same  magnitude of $C_p$ as stable solutions and maintain $S{=}A/4$. This would translate into needing a much larger entropy than stable solutions,  explaining the super--entropicity phenomenon. 

In conclusion,  this Letter  describes a new quantum instability in extended gravitational thermodynamics, available when the cosmological constant can vary.  The instability results in a runaway evaporation of the solution. The diagnostic tool is the sign of $C_V$. It is also conjectured that the instability is intimately connected to the property of being super--entropic.  This was proven here for the charged BTZ black hole, and successfully tested for the large family of super--entropic solutions presented in ref.~\cite{Hennigar:2014cfa}. Some partial arguments forging a general connection between the two phenomena were presented. It would be interesting to test other solutions, and of course to find a proof of the conjecture. It is worth remarking that if a version of this instability is present for $\Lambda{>}0$, it could have astrophysical and cosmological  relevance if models of dynamical $\Lambda$ were to find application in those contexts.
 
 \medskip
 
 \begin{acknowledgments}
CVJ  thanks the  US Department of Energy for support under grant  DE-SC 0011687, and Amelia for her support and patience.    
\end{acknowledgments}

\bibliographystyle{apsrev4-1}
\bibliography{unstable_btz}

\end{document}